\documentstyle[epsfig]{aipproc}
\begin{document}
\begin{flushright}
YNU-HEPTh-00-103 \\
KUCP-170 \\
October 2000
\end{flushright}
\vspace*{-1cm}

\title{Polarized and Unpolarized Structures\\
of the Virtual Photon\thanks{
Presented by T.~Uematsu at PHOTON2000, Ambleside, England, 
26st-31th August 2000.}}

\author{Ken Sasaki$^*$ and Tsuneo Uematsu$^{\dagger}$}
\address{$^*$Department of Physics, Faculty of Engineering, 
Yokohama National University, \\ Yokohama 240-8501, Japan\\
$^{\dagger}$Department of Fundamental Sciences, FIHS, 
Kyoto University, Kyoto 606-8501, Japan}

\maketitle

\vspace*{-0.8cm}

\begin{abstract}
We discuss the structure functions and the parton distributions
in the virtual photon target, both polarized and unpolarized, 
beyond the leading order in QCD. We study the 
factorization-scheme dependence of the parton distributions. 
\end{abstract}

\vspace*{-0.5cm}

\section*{Introduction}
As Maria Krawczyk remarked in her introductory talk on structure 
functions \cite{K}
, the virtual photon structure provides a unique test of QCD. 
In this talk I would like to discuss the polarized and unpolarized virtual
photon structures. But because of the limitation of the allocated time, 
I will mainly focus my talk on the polarized virtual photon structure. 

Recently there has been growing interest in the polarized 
photon structure functions. Especially, the 1st moment of a photon
structure function $g_1^\gamma$ has attracted much attention in connection with
its relevance for the axial anomaly, which has also played an important role 
in the QCD analysis of the nucleon spin structure functions. 
Now the information on the spin structure of the photon will be
obtained from the resolved photon process in polarized electron and
proton collision in the polarized version of the $ep$ collider.
More directly, the spin-dependent structure function of the photon
can be measured by the polarized $e^+ e^-$ collision in the future
linear colliders. 

Here we investigate two-photon process (Figure 1) with the kinematical region
where the mass squared of the probe photon ($Q^2$) is much larger than 
that of the target photon ($P^2$) which is in turn much bigger than the
$\Lambda^2$, the QCD scale parameter squared. The advantage for studying
the virtual photon target is that we can calculate whole structure functions
up to next-leading-order (NLO), in contrast to the real photon target where
there remain uncalculable non-perturbative pieces. This is true for
summing up the QCD logarithmic terms due to twist-2 operators corresponding
to the QCD parton picture. Here we neglect all the power corrections arising 
from the higher-twist effects and target mass effects of the form 
$(P^2/Q^2)^k$ ($k=1,2,\cdots$).  Some non-perturbative effects like gluon 
condensations reside in the higher-twist effects.
Our aim here is to study the polarized
virtual photon structure function $g_1^\gamma(x,Q^2,P^2)$
at the same level of unpolarized structure function $F_2^\gamma(x,Q^2,P^2)$
(For the experimental status, see \cite{SR}).
We can also investigate the parton distributions inside the polarized 
virtual photon. 
As we will see, the spin structure of the polarized virtual photon
would offer a good testing ground for factorization scheme dependence
of the parton distribution functions.

\begin{figure}[h] % fig 1
\leftline{\hspace*{0.3in}\epsfig{file=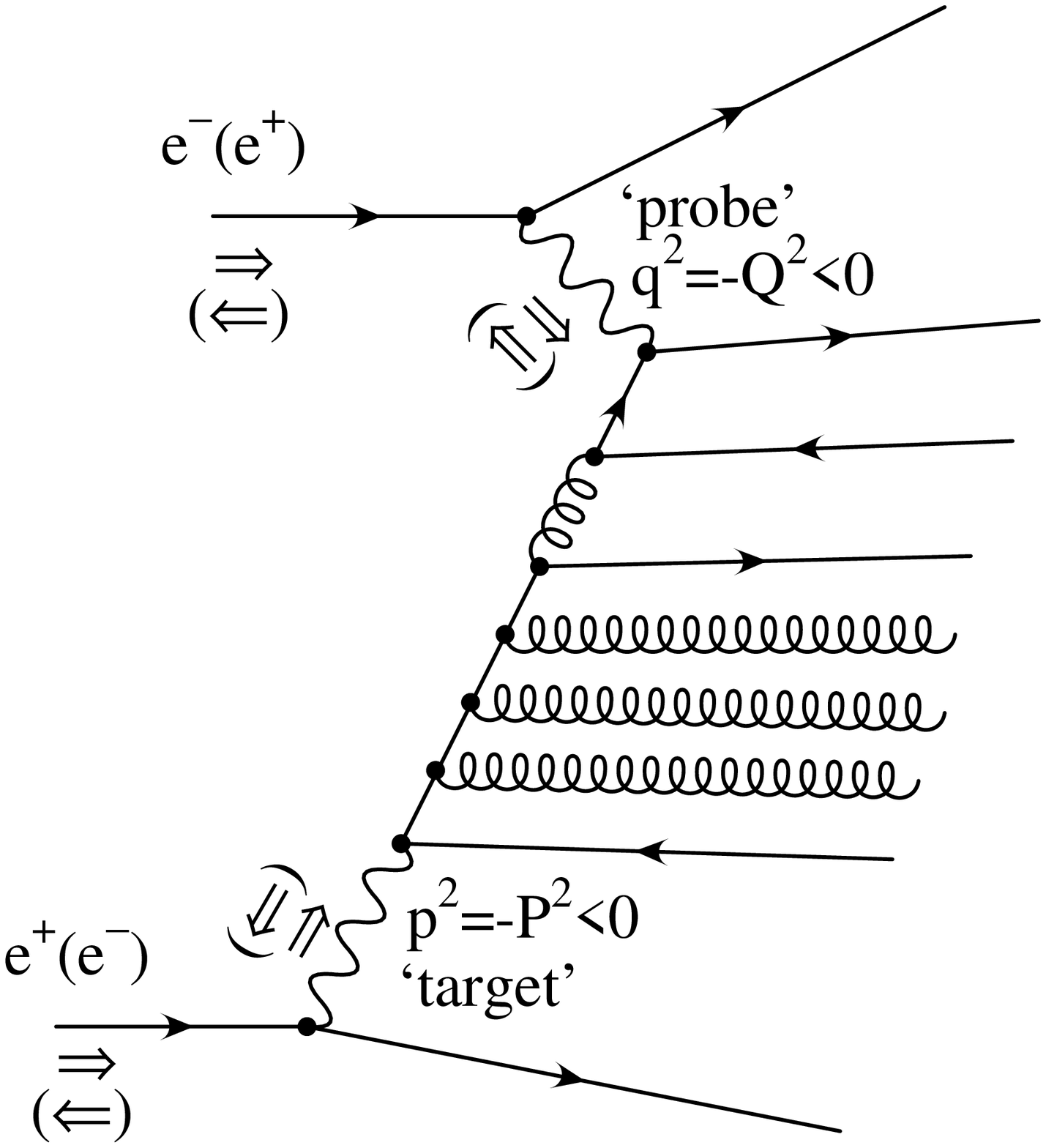,height=2.0in,width=2.0in}}
\vspace*{-3.2in}\hspace*{2.8in}
{\epsfig{file=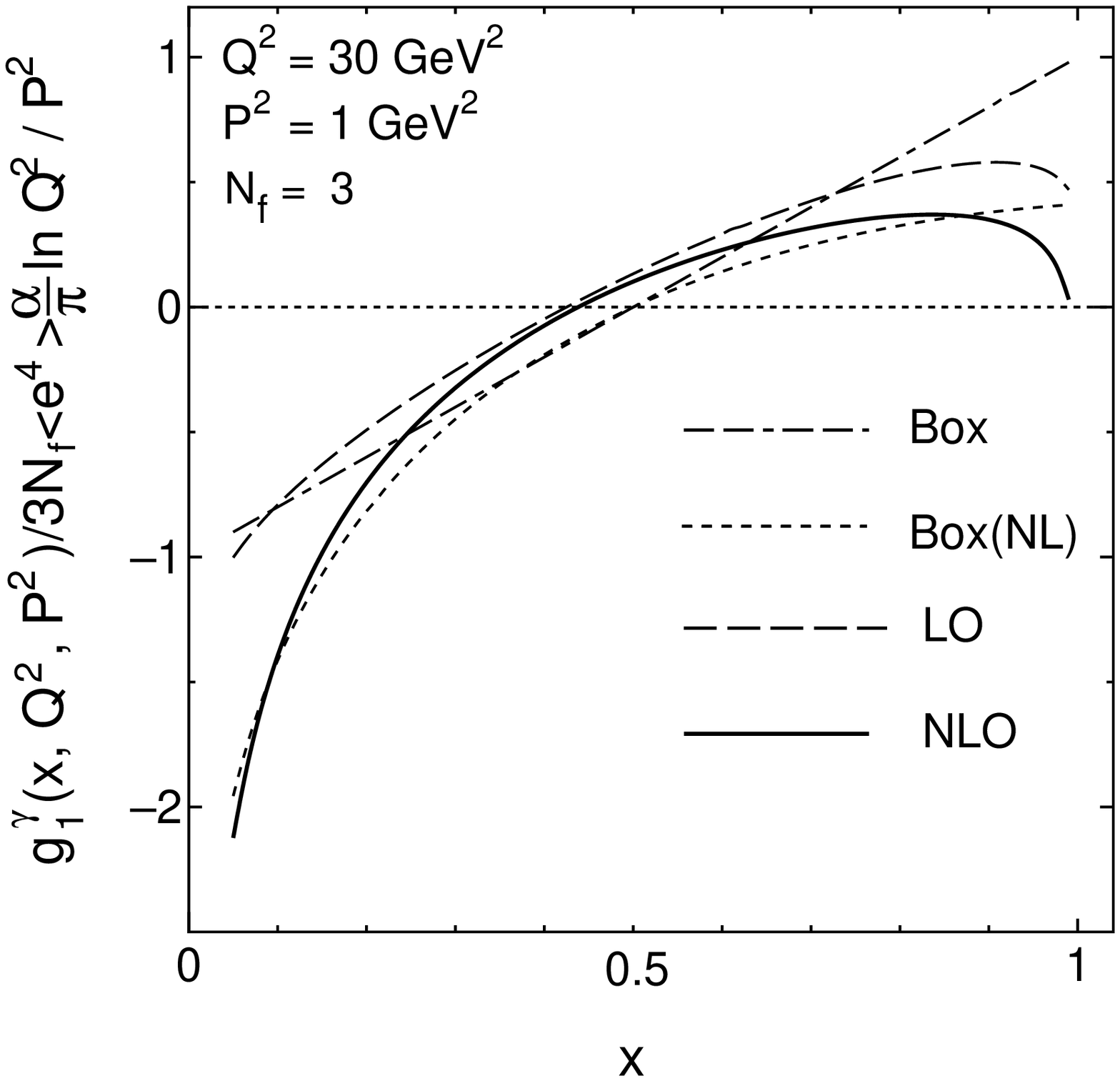,height=4.5in,width=3.0in}}
\vspace*{-1.0in}
\caption{Two-photon process in polarized $e^+e^-$ collision 
for $\Lambda^2 \ll P^2 \ll Q^2$ and the polarized photon structure
function $g_1^\gamma(x,Q^2,P^2)$ for $Q^2=30$ GeV$^2$ and $P^2=1$ GeV$^2$
with $N_f=3$. LO, NLO and Box (NL) denote QCD LO, NLO and Box-diagram
(non-leading) results.}
\label{fig1}
\end{figure}
\vspace*{-0.2in}

\section*{PQCD Calculation}
We can apply the same framework used in the analysis of nucleon spin
structure functions, namely the operator product expansion (OPE) supplemented
by the renormalization group (RG) method or equivalently 
DGLAP type parton evolution equations.   
The NLO calculation has become possible
since the two-loop anomalous dimensions of the quark and gluon operators
in OPE or equivalently two-loop parton splitting functions were calculated
by two groups \cite{MvN,V}
The $n$-th moment of $g_1^\gamma(x,Q^2,P^2)$ for the kinematical region:
$$
\Lambda^2 \ll P^2 \ll Q^2
$$
is given by
\begin{eqnarray}
%&&\int_0^1 dx x^{n-1}g_1^\gamma(x,Q^2,P^2)  \nonumber\\
&&\int_0^1 dx x^{n-1}g_1^\gamma(x,Q^2,P^2)  
=\frac{\alpha}{4\pi}\frac{1}{2\beta_0}
\left[\sum_{i=+,-,NS}
L_i^n
\frac{4\pi}{\alpha_s(Q^2)}
\left\{1-\left(\frac{\alpha_s(Q^2)}{\alpha_s(P^2)}\right)^{\lambda_i^n/2\beta_0
+1}\right\}\right.\nonumber\\
&&\left.\hspace{0.2cm}+\sum_{i=+,-,NS}{\cal A}_i^n\left\{1-\left(\frac{\alpha_s(Q^2)}{\alpha_s(P^2)}\right)^{\lambda_i^n/2\beta_0}\right\}
+\sum_{i=+,-,NS}{\cal B}_i^n
\left\{1-\left(\frac{\alpha_s(Q^2)}{\alpha_s(P^2)}\right)^{\lambda_i^n/2\beta_0+1}\right\}\right.\nonumber\\
&&\left.\hspace{9cm}+{\cal C}^n +{\cal O}(\alpha_s) \ \right.\Bigr]
\label{master}
\end{eqnarray}
where $L_i^n$, ${\cal A}_i^n$, ${\cal B}_i^n$ and ${\cal C}^n$
are computed from the 1- and 2-loop anomalous dimensions as well as
from 1-loop coefficient functions. $\lambda_i^n \ (i=+,-,NS)$
denote the eigenvalues of 1-loop anomalous dimension matrices.
$\alpha_s(Q^2)$ is the QCD running coupling constant. 
%(\ref{master})is our master equation.
In Figure 1 we have shown the $g_1^\gamma(x,Q^2,P^2)$  
evaluated from (\ref{master}) by inverse Mellin transform for
$Q^2=30$ GeV$^2$ and $P^2=1$ GeV$^2$ with $N_f=3$ \cite{SU3}.\\
Note that the same formula with different coefficients, 
$L_i^n$, ${\cal A}_i^n$, ${\cal B}_i^n$, ${\cal C}^n$ and 
$\lambda_i^n \ (i=+,-,NS)$ holds for the unpolarized structure function 
$F_2^\gamma(x,Q^2,P^2)$ \cite{UW}.

\section*{Sum Rule}

For a real photon target ($P^2=0$), Bass, Brodsky and Schmidt
have shown that  
the 1st moment of $g_1^\gamma(x,Q^2)$
vanishes to all orders of $\alpha_s(Q^2)$ in QCD
\cite{BBS}:
\begin{equation}
\int_0^1 dx g_1^\gamma(x,Q^2)=0.
\end{equation}
Now the question is what about the $n=1$ moment of the virtual photon case.
Here we note that the eigenvalues of one-loop anomalous dimension 
matrix 
%${\hat\gamma}_{n=1}^{(0)}$:
are $\lambda_+^{n=1}=0, \lambda_{-}^{n=1}=-2\beta_0$. 
Taking $n\rightarrow 1$ limit of 
(\ref{master})
the first three terms vanish.
Denoting 
$e_i$, the $i$-th quark charge and $N_f$, the number of active flavors,
we have
\begin{equation}
\int_0^1 dx g_1^\gamma(x,Q^2,P^2)=
-\frac{3\alpha}{\pi}\sum_{i=1}^{N_f}{e_i}^4
+{\cal O}(\alpha_s)
\end{equation}
We can go a step further to ${\cal O}(\alpha_s)$ QCD corrections
which turn out to be \cite{SU}:
\begin{eqnarray}
\int_0^1dx g_1^\gamma(x,Q^2,P^2)
&=&-\frac{3\alpha}{\pi}
\left[\sum_{i=1}^{N_f}e_i^4\left(1-\frac{\alpha_s(Q^2)}{\pi}\right)
-\frac{2}{\beta_0}(\sum_{i=1}^{N_f}e_i^2)^2\left(
\frac{\alpha_s(P^2)}{\pi}-\frac{\alpha_s(Q^2)}{\pi}\right)\right]\nonumber\\
&&\hspace{2cm}+{\cal O}(\alpha_s^2).
\end{eqnarray}
This result coincides with the one obtained by Narison, Shore and 
Veneziano \cite{NSV}, apart from the overall sign for the definition
of $g_1^\gamma$.

\section*{Parton Distributions}
\subsection*{Spin-dependent parton distributions}

Factorization theorem tells us that the physically observable
quantities like cross sections or structure functions can be
factored into the long-distance part (distribution function) and
short-distance part (coefficient function). Thus the polarized 
photon structure function can be written schematically as
\begin{equation}
g_1^\gamma=\Delta\vec{q}^\gamma \ \otimes\  \Delta\vec{C}^\gamma
\label{factorization}
\end{equation}
where spin-dependent parton distributions $\Delta\vec{q}$:
\begin{equation}
\Delta\vec{q}^\gamma(x,Q^2,P^2)=(\Delta q_S^\gamma,\Delta
 G^\gamma, \Delta q_{NS}^\gamma, \Delta \Gamma^\gamma)
\end{equation}
are polarized flavor-singlet quark, gluon, non-singlet quark 
and photon distribution functions in the virtual photon
(we put the symbol $\Delta$ for polarized quantities),
and  
\begin{equation}
\Delta\vec{C}^\gamma=\pmatrix
{\Delta C_S^\gamma \cr \Delta C_G^\gamma \cr \Delta C_{NS}^\gamma \cr
\Delta C_\gamma^\gamma}
\end{equation}
are the corresponding coefficient functions.   
The same relation holds for unpolarized structure function $F_2^\gamma$
in terms of unpolarized parton distributions $\vec{q}$ 
and unpolarized coefficient functions $\vec{C}^\gamma$.
In the leading order in QED coupling $\alpha=\frac{e^2}{4\pi}$, the
photon distribution function can be taken as 
$\Delta \Gamma^\gamma (x,Q^2,P^2)=\delta(1-x)$. Therefore
we have the following inhomogeneous DGLAP evolution equation for
$\Delta\mbox{\boldmath $q$}^\gamma=(\Delta q_S^\gamma,\Delta
 G^\gamma, \Delta q_{NS}^\gamma)$:
\begin{equation}
\frac{d\Delta\mbox{\boldmath $q$}^\gamma(x,Q^2,P^2)}{d\ln Q^2}=
\Delta\mbox{\boldmath $K$}(x,Q^2)+\int_x^1\frac{dy}{y}
\Delta\mbox{\boldmath $q$}^\gamma
(y,Q^2,P^2)\times \Delta P(\frac{x}{y},Q^2)\label{DGLAP}
\end{equation}
where $\Delta\mbox{\boldmath $K$}(x,Q^2)$ is the splitting function of 
the photon into quark and gluon, whereas $\Delta P({x}/{y},Q^2)$ is 
the 3$\times$3 splitting function matrix.

\subsection*{Factorization Scheme Dependence}
The solution to the DGLAP evolution equation can be given by
\begin{equation}
\Delta\vec{q}^\gamma(t)=\Delta\vec{q}^{\gamma(0)}(t)
+\Delta\vec{q}^{\gamma(1)}(t),\quad 
t \equiv \frac{2}{\beta_0}\ln \frac{\alpha_s(P^2)}{\alpha_s(Q^2)}
\end{equation}
where the first (second) term corresponds to LO (NLO) approximation.
The initial condition we impose is the following,
\begin{equation}
\Delta\vec{q}^{\gamma(0)}(0)=0, \quad 
\Delta\vec{q}^{\gamma(1)}(0)=\frac{\alpha}{4\pi}\vec{A}_n
\end{equation}
where $\vec{A}_n$ is the constant which depends on the factorization
scheme to be used. 
Or equivalently in the language of OPE, this constant appears
as a finite matrix element of the operators, $\vec{O}_n$ renormalized at
$\mu^2=P^2$  
between the photon states:
\begin{equation}
\langle \gamma (p) \mid \vec{O}_n (\mu) \mid \gamma (p) 
\rangle|_{\mu^2=P^2} =\frac{\alpha}{4\pi}\vec{A}_n
\end{equation}
This scheme dependence arises from the freedom of multiplying the arbitrary
finite renormalization constant $Z_a$ and its inverse $Z_a^{-1}$
in the $n$-th moment of (\ref{factorization}):
\begin{eqnarray}
g_1^\gamma(n,Q^2,P^2)=\Delta\vec{q}^\gamma\cdot\Delta\vec{C}^\gamma
=\Delta\vec{q}^\gamma Z_a \cdot 
Z_a^{-1}\Delta\vec{C}^\gamma
=\Delta\vec{q}^\gamma|_a \cdot \Delta\vec{C}^\gamma|_a
\end{eqnarray}
where the resulting $\Delta\vec{q}^\gamma|_a$ and $\Delta\vec{C}^\gamma|_a$
are the distribution function and the coefficient function in the $a$-scheme.
The explicit expressions for the $n$-th moment of
the parton distributions can be found in ref.\cite{SU3}.

\subsection*{Transformation from $\overline{\rm MS}$ to $a$-scheme}
Under the transformation from one factorization scheme to another, the 
coefficient functions as well as anomalous dimensions will change.
Of course when they are combined together, we get the factorization-scheme
independent structure function $g_1^\gamma$. Since $\overline{\rm MS}$ 
is the only scheme in which both 1-loop coefficient functions and 2-loop 
anomalous dimensions are actually computed, we study the transformation
rule from the $\overline{\rm MS}$ to a new factorization scheme-$a$.
We have considered the several different factorization schemes;
1) chirally invariant (CI) scheme, 2) Adler-Bardeen (AB) scheme, 3) off-shell
 (OS) scheme, 4) Altarelli-Ross (AR) scheme and 5) ${\rm DIS}_\gamma$ scheme.
(For the detailed description of each factorization scheme see \cite{SU3}.)
The transformation rule for the singlet-quark coefficient function, for
example, is given by
\begin{equation}
\Delta C_{S,~a}^{\gamma,~ n}=
\Delta C_{S,~\overline{\rm MS}}^{\gamma,~ n}~
-\langle e^2\rangle\frac{\alpha_s}{2\pi}~\Delta w(n,a)
\end{equation}
where $\langle e^2\rangle=\sum_i e_i^2/N_f$ and 
$\Delta w(n,a)$ is the transformation functions, 
the explicit expressions for which 
as well as other coefficient functions together with
the similar transformation rules for 2-loop
anomalous dimensions are given in ref.\cite{SU3}. 

The prescriptions to treat the axial anomaly are different from
scheme to scheme. For example, the axial anomaly resides in the
quark distribution in  $\overline{\rm MS}$ schme, whereas it exists
in the gluon and photon coefficient functions in the CI scheme.
These factorization schemes are also characterized by the behavior
of the parton distribution functions near $x=1$. We can study their
analytic behaviors by the large $n$ limit of their moments.
Here we also note that the gluon distribution function is 
factorization-scheme independent in the class of factorization 
schemes considered here. 
By performing the inverse Mellin transform of the moments, 
the parton distributions as functions of $x$ are reproduced numerically.
We present our results for singlet-quark for various schemes and gluon
in Figure 2. Note that the real photon's $g_1^\gamma$ was studied in
\cite{S,SV}, which are consistent with present analysis.

The similar scheme dependence of the parton
distributions inside the unpolarized virtual photon was studied
for the $\overline{\rm MS}$, OS and ${\rm DIS}_\gamma$ schemes
\cite{SU4}.

\begin{figure}[h] % fig 3-4
\leftline{\hspace*{0.1in}\epsfig{file=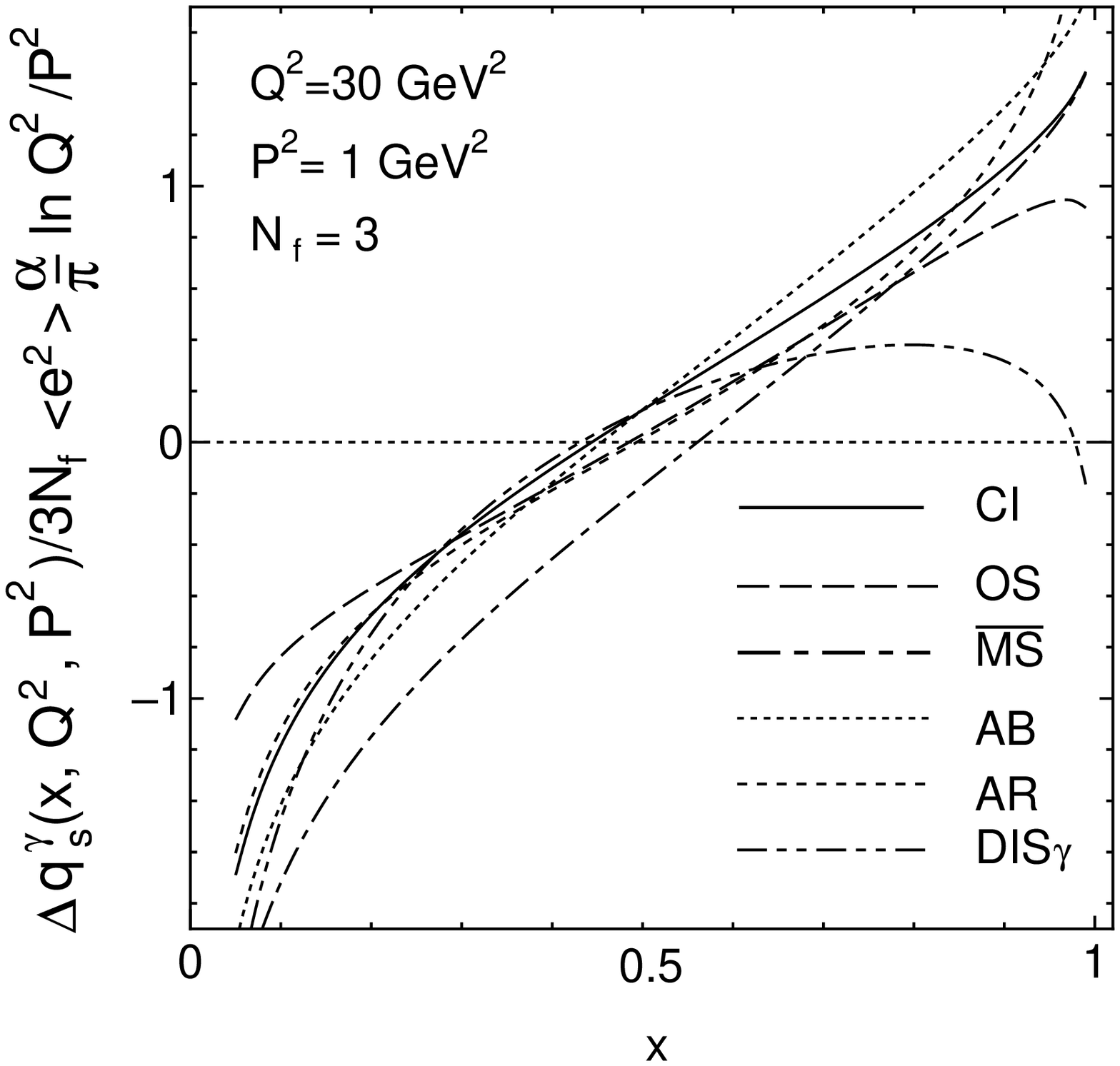,height=4.5in,width=3.0in}}
\vspace*{-4.5in}\hspace*{2.9in}
{\epsfig{file=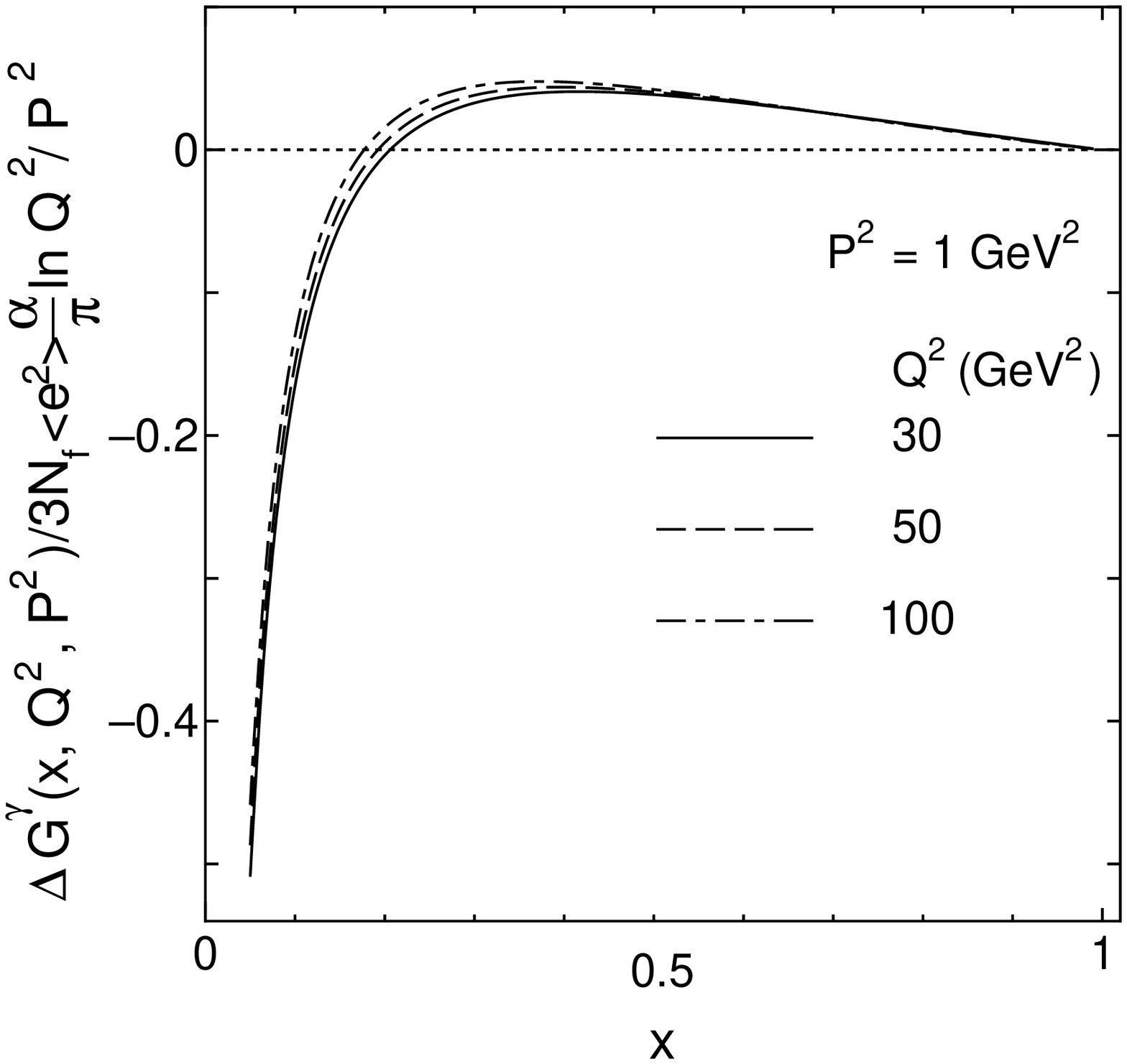,height=4.5in,width=3.0in}}
\vspace*{-1.0in}
\caption{Polarized singlet-quark distribution function 
$\Delta q_S^\gamma(x,Q^2,P^2)$ in the several factorization schemes
and the polarized gluon distribution $\Delta G^\gamma(x,Q^2,P^2)$
for $Q^2=30$ GeV$^2$ and $P^2=1$ GeV$^2$ with $N_f=3$ }
\label{fig3}
\end{figure}

\section*{Concluding Remarks}

We have studied the virtual photon's spin structure functions,
$g_1^\gamma(x,Q^2,P^2)$ and the polarized parton distributions
for the kinematical region $\Lambda^2 \ll P^2 
\ll Q^2$, which are perturbatively calculable up to the NLO in QCD.
The first moment of $g_1^\gamma$ is non-vanishing in contrast to the 
real photon case, where we have vanishing sum rule.
NLO QCD corrections are significant at large $x$ as well as at low $x$.
We also studied
factorization-scheme dependence of parton distribution functions.

Future subjects to be studied are as follows. First of all we should
understand how the transition occurs from vanishing 1st moment for real 
photon ($P^2=0$) to non-vanishing one for virtual photon ($P^2 \gg \Lambda^2$).
Secondly, another structure function $g_2^\gamma(x,Q^2,P^2)$ is yet to be
computed where we also have twist-3 contribution. Furthermore,
the power corrections due to target mass effects and higher-twist effects
should be investigated.
More reliable treatment for small-$x$ behaviors of polarized p.d.f. 
should be studied in the framework of BFKL like approach.


\vspace*{-0.4cm}


\begin{references}
\bibitem{K} Krawczyk, M., these proceedings.
\bibitem{SR} S{\"o}ldner-Rembold, S., these proceedings.
\bibitem{MvN} Mertig, R., and van Neerven, W. L., 
{\it Z.\ Phys.}\ {\bf C70}, 637 (1996). 
\bibitem{V} Vogelsang, W., {\it Phys. Rev.}\  {\bf D54}, 2023 (1996).
\bibitem{UW} Uematsu, T., and Walsh, T. F., {\it Nucl. Phys.}\
{\bf B199}, 93 (1982).
\bibitem{BBS} Bass, S. D., Brodsky, S. J., and Schmidt, I., Phys. Lett.
{\bf B437}, 417 (1998).
\bibitem{SU} Sasaki, K., and Uematsu, T., {\it Phys. Rev.}\ {\bf D59},
114011 (1999).
\bibitem{NSV} Narison, S., Shore, G. M., and Veneziano, G., 
{\it Nucl. Phys.}\ {\bf B391}, 69 (1993).
\bibitem{SU2} Sasaki, K., and Uematsu, T., {\it Phys. Lett.}\ {\bf B473},
309 (2000).
\bibitem{SU3} Sasaki, K., and Uematsu, T., hep-ph/0007055 (2000).
\bibitem{S} Sasaki, K., {\it Phys. Rev.} {\bf D22}, 2143 (1980); 
{\it Prog. Theor. Phys. Suppl.}\ {\bf 77}, 197 (1983).
\bibitem{SV} Stratmann, M., and Vogelsang, W., 
{\it Phys. Lett.}\ {\bf B386}, 370 (1996).
\bibitem{SU4} Sasaki, K., and Uematsu, T., 
{\it Nucl. Phys. B (Proc. Suppl.)}\ {\bf 89}, 162 (2000).
\end{references}
\end{document}